\documentclass[aps,prl,twocolumn,showpacs,superscriptaddress,nofootinbib]{revtex4}

\usepackage{amsmath}    
\usepackage{graphicx}   

\usepackage{xspace}
\usepackage{units}

\usepackage{hyperref}

\newcommand{\minerva}{MINERvA\xspace}
\newcommand{\minos}{MINOS\xspace}

\newcommand{\numu}{\ensuremath{\nu_{\mu}}\xspace}

\newcommand{\Qsqqe}{\ensuremath{Q^{2}_{QE}}\xspace}
\newcommand{\dsdq}{\ensuremath{d\sigma/dQ^2}\xspace}

\newcommand{\GeV}{\ensuremath{\mbox{GeV}}\xspace}

\newcommand{\GeVcc}{\ensuremath{\mbox{GeV}/c^2}\xspace}

\newcommand{\footnoteremember}[2]{\footnote{#2}\newcounter{#1}\setcounter{#1}{\value{footnote}}}
\newcommand{\footnoterecall}[1]{\footnotemark[\value{#1}]}

\newcommand{\sizecheck}{0} 
\newcommand{\PRLsupp}{0}   
\ifnum\PRLsupp=0
  
\else
  
\fi

\newif\ifpdf
\ifx\pdfoutput\undefined
   \pdffalse
\else
   \pdfoutput=1
   \pdftrue
\fi
\ifpdf
   \usepackage{graphicx}
   \usepackage{epstopdf}
\else
   \usepackage{graphicx}
\fi

\begin{document}
\title{Measurement of Muon Neutrino Quasi-Elastic Scattering \\ on a Hydrocarbon Target at $E_{\nu} \sim$ 3.5~GeV}


\newcommand{\deceased}{Deceased}

\newcommand{\wroclaw}{Institute of Theoretical Physics, Wroc\l aw University, Wroc\l aw, Poland}    
\newcommand{\Rutgers}{Rutgers, The State University of New Jersey, Piscataway, New Jersey 08854, USA}
\newcommand{\Hampton}{Hampton University, Dept. of Physics, Hampton, VA 23668, USA}
\newcommand{\Dortmund}{Institute of Physics, Dortmund University, 44221, Germany }
\newcommand{\Otterbein}{Department of Physics, Otterbein University, 1 South Grove Street, Westerville, OH, 43081 USA}
\newcommand{\JMU}{James Madison University, Harrisonburg, Virginia 22807, USA}
\newcommand{\Florida}{University of Florida, Department of Physics, Gainesville, FL 32611}
\newcommand{\UCIrvine}{Department of Physics and Astronomy, University of California, Irvine, Irvine, California 92697-4575, USA}
\newcommand{\CBPF}{Centro Brasileiro de Pesquisas F\'{i}sicas, Rua Dr. Xavier Sigaud 150, Urca, Rio de Janeiro, RJ, 22290-180, Brazil}
\newcommand{\PUCP}{Secci\'{o}n F\'{i}sica, Departamento de Ciencias, Pontificia Universidad Cat\'{o}lica del Per\'{u}, Apartado 1761, Lima, Per\'{u}}
\newcommand{\INRM}{Institute for Nuclear Research of the Russian Academy of Sciences, 117312 Moscow, Russia}
\newcommand{\Jlab}{Jefferson Lab, 12000 Jefferson Avenue, Newport News, VA 23606, USA}
\newcommand{\Pittsburgh}{Department of Physics and Astronomy, University of Pittsburgh, Pittsburgh, Pennsylvania 15260, USA}
\newcommand{\Guanajuato}{Campus Le\'{o}n y Campus Guanajuato, Universidad de Guanajuato, Lascurain de Retana No. 5, Col. Centro. Guanajuato 36000, Guanajuato M\'{e}xico.}
\newcommand{\Athens}{Department of Physics, University of Athens, GR-15771 Athens, Greece}
\newcommand{\Tufts}{Physics Department, Tufts University, Medford, Massachusetts 02155, USA}
\newcommand{\WM}{Department of Physics, College of William \& Mary, Williamsburg, Virginia 23187, USA}
\newcommand{\FNAL}{Fermi National Accelerator Laboratory, Batavia, Illinois 60510, USA}
\newcommand{\Purdue}{Department of Chemistry and Physics, Purdue University Calumet, Hammond, Indiana 46323, USA}
\newcommand{\MCLA}{Massachusetts College of Liberal Arts, 375 Church Street, North Adams, MA 01247}
\newcommand{\UMD}{Department of Physics, University of Minnesota -- Duluth, Duluth, Minnesota 55812, USA}
\newcommand{\Northwestern}{Northwestern University, Evanston, Illinois 60208}
\newcommand{\UNI}{Universidad Nacional de Ingenier\'{i}a, Apartado 31139, Lima, Per\'{u}}
\newcommand{\Rochester}{University of Rochester, Rochester, New York 14610 USA}
\newcommand{\Austin}{Department of Physics, University of Texas, 1 University Station, Austin, Texas 78712, USA}
\newcommand{\USM}{Departamento de F\'{i}sica, Universidad T\'{e}cnica Federico Santa Mar\'{i}a, Avda. Espa\~{n}a 1680 Casilla 110-V, Valpara\'{i}so, Chile}
\newcommand{\Geneva}{University of Geneva, Geneva, Switzerland}
\newcommand{\Chicago}{Enrico Fermi Institute, University of Chicago, Chicago, IL 60637 USA}
\newcommand{\keppelThanks}{\thanks{now at the Thomas Jefferson National Accelerator Facility, Newport News, VA 23606 USA}}
\newcommand{\giulianoThanks}{\thanks{now at Vrije Universiteit Brussel, Pleinlaan 2, B-1050 Brussels, Belgium}}
\newcommand{\LazaThanks}{\thanks{also at Department of Physics, University of Antananarivo, Madagascar}}
\newcommand{\schulteThanks}{\thanks{now at Temple University, Philadelphia, Pennsylvania 19122, USA}}
\newcommand{\jwaldingThanks}{\thanks{now at Dept. Physics, Royal Holloway, University of London, UK}}

\author{G.A.~Fiorentini}                  \affiliation{\CBPF}
\author{D.W.~Schmitz}                     \affiliation{\Chicago}  \affiliation{\FNAL}
\author{P.A.~Rodrigues}                   \affiliation{\Rochester}
\author{L.~Aliaga}                        \affiliation{\WM}  \affiliation{\PUCP}
\author{O.~Altinok}                       \affiliation{\Tufts}
\author{B.~Baldin}                        \affiliation{\FNAL}
\author{A.~Baumbaugh}                     \affiliation{\FNAL}
\author{A.~Bodek}                         \affiliation{\Rochester}
\author{D.~Boehnlein}                     \affiliation{\FNAL}
\author{S.~Boyd}                          \affiliation{\Pittsburgh}
% ...
\collaboration{The \minerva Collaboration}\ \noaffiliation
\date{\today}
\begin{abstract}
We report a study of \numu charged-current quasi-elastic events 
in the segmented scintillator inner tracker of the \minerva 
experiment running in the NuMI neutrino beam at Fermilab. 
The events were selected by requiring a $\mu^-$ and low calorimetric recoil 
energy separated from the interaction vertex. We measure the flux-averaged 
differential cross-section, \dsdq, and study the low energy particle content 
of the final state.   Deviations are found between the measured \dsdq and 
the expectations of a model of independent nucleons in a 
relativistic Fermi gas.  We also observe an excess of energy near 
the vertex consistent with multiple protons in the final state.  

\end{abstract}

\ifnum\sizecheck=0
\maketitle
\fi

Charged-current neutrino quasi-elastic scattering,
$\nu_\mu n\to\mu^-p$, distinguishes neutrino flavor and is valuable for 
neutrino oscillation experiments at energies near $1$~GeV
where it is responsible for a large fraction of the total reaction
cross-section~\cite{AguilarArevalo:2010wv,AguilarArevalo:2008rc,Abe:2011sj,Ayres:2004js}. For free nucleons the scattering process may be described by the standard theory of weak interactions with the inclusion of nucleon form factors~\cite{LlewellynSmith:1971zm}.   
Electron scattering~\cite{Bradford:2006yz} and neutrino scattering on deuterium~\cite{Bodek:2007vi,Kuzmin:2007kr} determine the most important form factors with good precision~\cite{Day:2012gb}.
However, neutrino oscillation experiments typically use detectors made of
heavier nuclei such as carbon~\cite{AguilarArevalo:2008qa,Ayres:2004js}, 
oxygen~\cite{Abe:2011ks}, iron~\cite{Michael:2008bc},  
or argon~\cite{Chen:2007ae, LBNE} where interactions with nucleons are 
modified by the nuclear environment.
These effects are commonly modeled using a 
relativistic Fermi gas~\cite{Smith:1972xh,Bodek:1980ar} 
(RFG) description of the nucleus as quasi-free, independent
nucleons with Fermi motion in a uniform binding potential.
Neutrino interaction generators~\cite{Andreopoulos201087,Hayato:2009zz,Golan:2012wx,Buss:2011mx,Lalakulich:2012gm} additionally simulate interactions of final state hadrons inside the target nucleus. The MiniBooNE experiment recently observed that this 
prescription, utilizing the free deuterium value for the axial form factor, does not accurately describe their
measurements of quasi-elastic scattering of neutrinos and
antineutrinos on a hydrocarbon target~\cite{AguilarArevalo:2007ab,AguilarArevalo:2013hm}.

The RFG approach may be supplemented by accounting for correlations between 
nucleons within the nucleus. Evidence for these correlations 
has been observed in electron-nucleus scattering~\cite{Subedi:2008zz}.
Processes that produce multiple final state nucleons are thought to lead to enhancements in the cross-section~\cite{Carlson:2001mp,Shen:2012xz,Bodek:2011ps}.
These contributions are modeled using
different approaches~\cite{Martini:2010ex,Martini:2013sha,Nieves:2005rq} which
produce qualitatively similar though not quantitatively identical results.
The RFG model may also be replaced by an alternate spectral 
function (SF) model that calculates the joint probability distribution of 
scattering off a nucleon of given momentum and binding energy 
inside a nucleus~\cite{Benhar:1994hw}.
These nuclear effects may be significant for oscillation experiments 
seeking to measure the neutrino mass hierarchy 
and CP violation~\cite{Nieves:2012yz,Lalakulich:2012hs,Martini:2012uc}.

In this Letter we report the first study of muon neutrino
quasi-elastic interactions at energies between $1.5$ and $10$~GeV from the \minerva experiment, which uses a
finely segmented scintillator detector at Fermilab to measure muon
neutrino interactions on 
nuclear targets.  The analysis technique is similar to the one
employed to study the antineutrino reaction~\cite{nubarprl}.  The signal
has a $\mu^-$ in the final state along with one or more nucleons
(typically with a leading proton), and no mesons.  We reject events in
which mesons are produced by requiring that the hadronic system
recoiling against the muon has a low energy. That energy is measured
in two spatial regions. The {\it vertex energy} region corresponds to
a sphere around the vertex with a radius sufficient to contain a
proton (pion) with 225 (100)~MeV kinetic energy. This region is
sensitive to low energy protons which could arise from correlations
among nucleons in the initial state or final state interactions of the outgoing
hadron inside the target nucleus~\cite{Sobczyk:2012ms}. 
We do not use the vertex energy in
the event selection but study it for evidence of multi-nucleon
processes.  The {\it recoil energy} region includes energy depositions
outside of the vertex region and is sensitive to pions and higher
energy nucleons. We use the recoil energy to estimate and remove
inelastic backgrounds.


The \minerva detector was exposed to the NuMI neutrino beam at Fermilab, configured for this analysis to produce a beam consisting of $>95\%$ $\nu_\mu$ at the peak energy of 3~GeV. The neutrino flux is predicted 
using a Geant4-based simulation tuned to hadron production 
data~\cite{Alt:2006fr} as described in Ref.~\cite{nubarprl}.
This analysis uses data taken between March and July 2010 with
$9.42\times 10^{19}$ protons on target.

The \minerva detector consists of a fine-grained scintillator tracker
surrounded by electromagnetic and hadronic 
calorimeters\footnote{The \minerva scintillator tracking region is 95\% CH and 5\% other materials by weight.}~\cite{minerva_nim}. The detector enables 
reconstruction of the neutrino interaction point, 
the tracks of outgoing charged particles, 
and the calorimetric reconstruction of other particles produced in the interaction. 
\minerva is located \unit[2]{m} upstream of the 
MINOS near detector~\cite{Michael:2008bc}, which is used to 
reconstruct the momentum and charge of muons.
The hadronic energy scale is set
using data from through-going muons and a scaled down 
\minerva detector exposed to a hadron test beam~\cite{minerva_nim}.  
The detector's performance is simulated by a Geant4-based hit-level 
simulation and a readout model tuned to match the data~\cite{minerva_nim}.   
Event pile-up causes a decrease in the muon track reconstruction efficiency 
which we studied in both \minerva and \minos by projecting tracks 
found in one of the detectors to the other 
and measuring the misreconstruction rate. 
This resulted in a -9.1\% (-4.8\%) correction to the simulated 
efficiency for muons with momenta below (above) 3 GeV/c in MINOS.
Neutrino interactions in the detector are simulated using the 
GENIE neutrino event generator~\cite{Andreopoulos201087}.  Details of the cross section models and associated parameters are described in
Ref.~\cite{nubarprl}.  

Event reconstruction and selection for this analysis is nearly identical
to that used in the \minerva antineutrino quasi-elastic
measurement~\cite{nubarprl} with small modifications to account for
the likelihood of a leading proton in the final state instead of a neutron.  
We require events to have a $\mu^-$ originating in the 
\unit[5.57]{metric ton} fiducial volume and assign remaining clusters with 
energies $>\unit[1]{MeV}$ to the recoiling hadronic system. 
The aforementioned vertex region corresponds to a sphere with
$\unit[30]{g/cm^2}$ of material centered on the vertex. 
The recoil system outside the vertex region is required to have $\leq 2$ isolated groups of spatially contiguous energy depositions\footnote{Isolated energy depositions are created directly by the leading proton or by secondary hadronic interactions in the detector.}.

The neutrino energy and the square of the four momentum transferred to the
nucleus, $Q^2_{QE}$, are estimated from the muon momentum and 
angle using a quasi-elastic hypothesis, as in the antineutrino
analysis~\cite{nubarprl}.  The binding energy correction is taken to be 
\unit[+34]{MeV} instead of \unit[+30]{MeV} used in Ref.~\cite{nubarprl} due to
Coulomb corrections~\cite{Katori:2008zz}, 
and the proton and neutron masses are interchanged.

\begin{figure}[tp]
\centering
\includegraphics[width=\columnwidth]{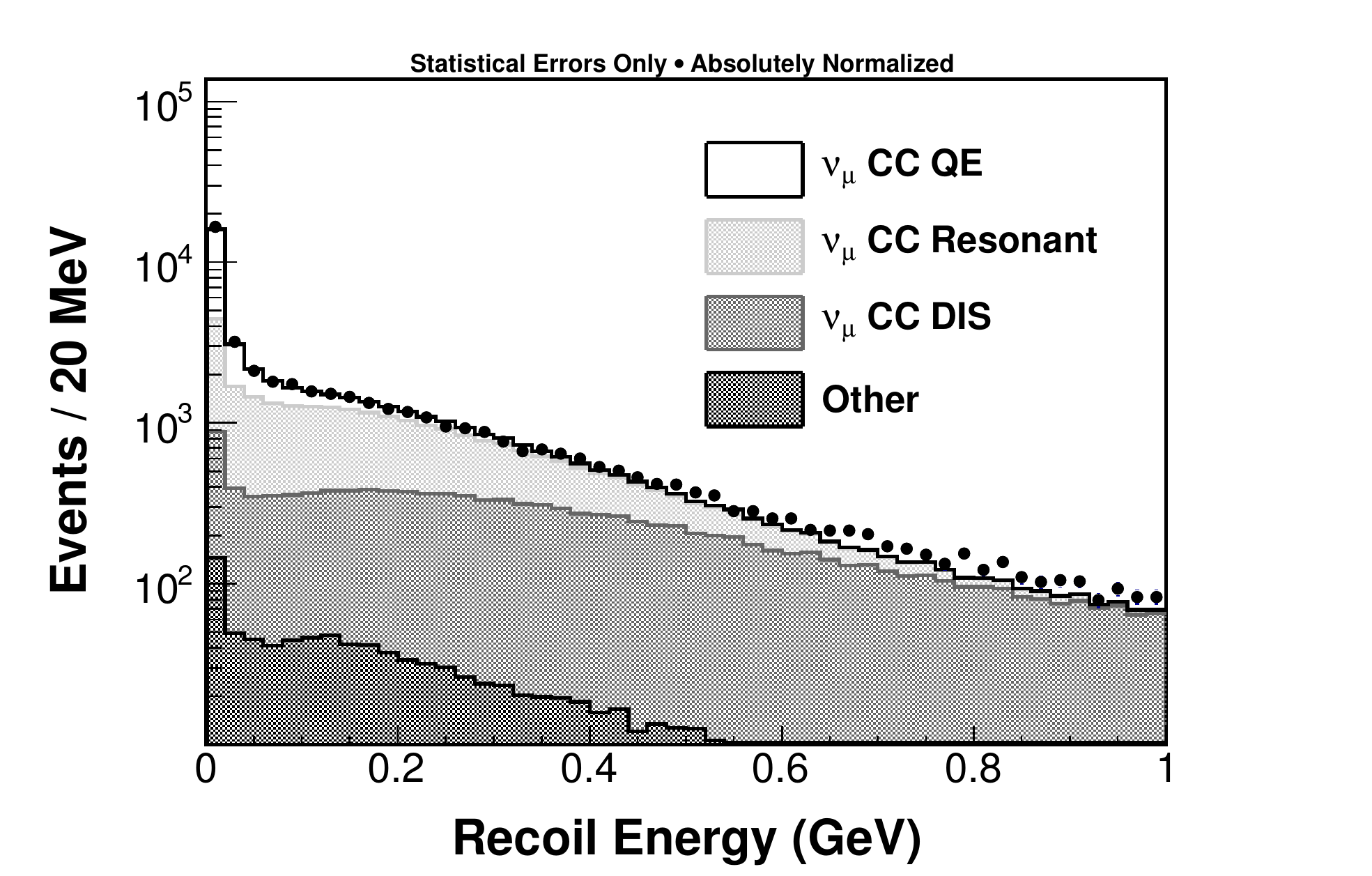}
\caption{\label{fig:recoil} The measured recoil energy distribution in the data (solid circles) and the predicted composition of signal and background. Backgrounds from baryon resonance production (light grey), continuum/deep-inelastic scattering (dark grey), and other sources (black), such as coherent pion production, are shown.  The fraction of signal before requiring low recoil energy is $0.29$.}
\end{figure}
\begin{figure}[tp]
\centering
\includegraphics[width=\columnwidth]{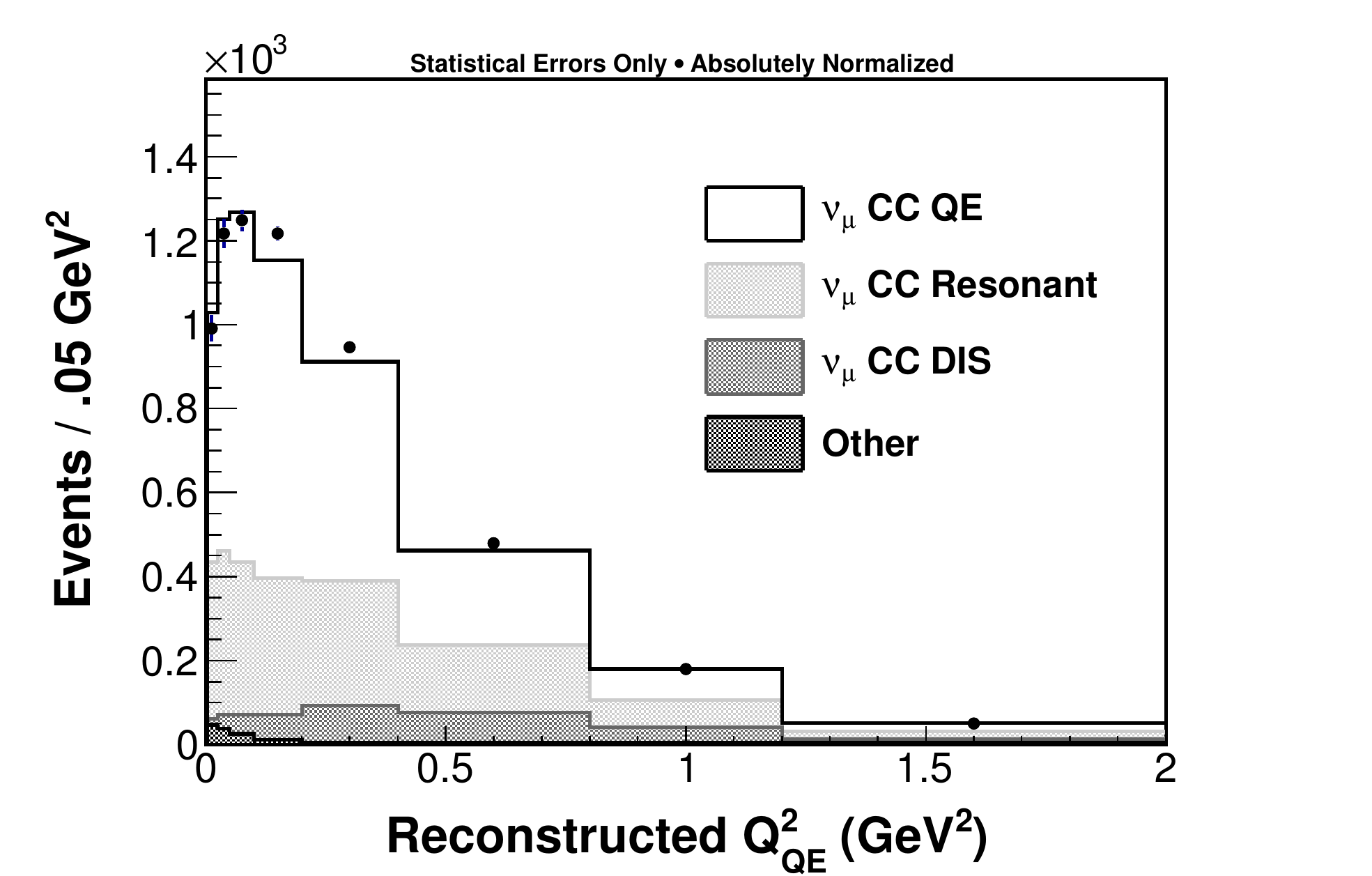}
\caption{The measured $Q^2_{QE}$ distributions in the data and the simulation, before correcting for detector resolution and acceptance.   The fraction of signal in this sample is $0.49$, and 47\% of signal events in our fiducial volume pass all selections.}
\label{fig:qsq}
\end{figure}

Figure~\ref{fig:recoil} shows that the quasi-elastic signal
preferentially populates lower recoil energies.  However, since the proton's kinetic energy is $\approx Q^2/2M_\text{neutron}$ for quasi-elastic scattering, the recoil 
energy is expected to scale with the momentum transfer as the 
final state proton becomes increasingly energetic and escapes the vertex region.
We account for this by varying a cut on the maximum 
allowed recoil energy as a function of \Qsqqe to assure 
95\% signal efficiency in each  \Qsqqe bin.  

The background in each $Q^2_{QE}$ bin is estimated from the data by fitting the relative normalizations of signal and background recoil energy distributions whose shapes are taken from the simulation.   This procedure reduces the relative background prediction by 15\% below $Q^2_{QE}$ of 0.8~GeV$^2$ and 5\% between 0.8 and 2.0~GeV$^2$.  The purity of the resulting sample 
ranges from 65\% at low \Qsqqe to 40\% at higher \Qsqqe.
Figure \ref{fig:qsq} compares the \Qsqqe distribution 
of the 29,620 events which satisfy the selection criteria to 
the simulation after rescaling the background according to the fit.
The cross-section as a function of \Qsqqe is extracted by subtracting the backgrounds, correcting for detector resolution and acceptance, and dividing by the number of neutrons in the fiducial volume ($1.65 \pm 0.02 \times 10^{30}$) and by the flux, as described in Ref.~\cite{nubarprl}.  The total neutrino 
flux integrated between 1.5 and 10~GeV is estimated by the simulation to be 
$2.91\times 10^{-8}/$cm$^2$ per proton on target\footnoteremember{footsupp}{See Supplemental Material\ SuppLocation\ for the flux as a function of energy and for correlations of uncertainties among bins for the cross-section and shape measurement.}.

The same systematic uncertainties which affect the antineutrino
analysis~\cite{nubarprl} are evaluated in this analysis.  
Table~\ref{tab:systematics} shows a summary of
systematic uncertainties on $d\sigma/dQ^2_{QE}$. 
The largest uncertainties on the absolute cross-section 
come from the neutrino flux and the muon momentum scale.  
However, the flux uncertainty is largely independent of $Q^2_{QE}$ so comparisons of the shape of $d\sigma/dQ^2_{QE}$  to scattering model predictions are relatively insensitive to knowledge of the flux. 
The saturation of ionization ($dE/dx$), parameterized by Birk's law and characterized by a factor of $(1 + k_B\times dE/dx)^{-1}$, leads to a recoil reconstruction uncertainty; this uncertainty is negligible for the antineutrino $d\sigma/dQ^2_{QE}$ measurement but is important for the neutrino measurement.
By studying stopping proton tracks in the \minerva test beam detector
we estimate $k_B=\unit[0.13\pm0.04]{mm/MeV}$~\cite{minerva_nim}, and vary
the ionization accordingly in the simulation to propagate the uncertainty.


\begingroup
\squeezetable
\begin{table}
\begin{tabular}{cccccccc}
$Q^2_{QE}$ (GeV$^2$) & I & II & III & IV & V & VI & Total \\
\hline
$0.0 - 0.025$ & 0.06 & 0.04 & 0.02 & 0.04 & 0.09 & 0.03 & 0.13 \\
$0.025 - 0.05$ & 0.06 & 0.03 & 0.02 & 0.03 & 0.09 & 0.02 & 0.12 \\
$0.05 - 0.1$ & 0.06 & 0.03 & 0.02 & 0.03 & 0.09 & 0.02 & 0.12 \\
$0.1 - 0.2$ & 0.06 & 0.03 & 0.03 & 0.02 & 0.09 & 0.02 & 0.11 \\
$0.2 - 0.4$ & 0.05 & 0.02 & 0.03 & 0.03 & 0.09 & 0.01 & 0.11 \\
$0.4 - 0.8$ & 0.05 & 0.03 & 0.04 & 0.04 & 0.09 & 0.01 & 0.13 \\
$0.8 - 1.2$ & 0.08 & 0.07 & 0.07 & 0.15 & 0.09 & 0.02 & 0.22 \\
$1.2 - 2.0$ & 0.12 & 0.07 & 0.07 & 0.16 & 0.09 & 0.02 & 0.24 \\
\hline
\end{tabular}
\caption{
Fractional systematic uncertainties on $d\sigma/dQ^2_{QE}$ associated with (I) muon reconstruction, (II) recoil reconstruction, (III) neutrino interaction models, (IV) final state interactions, (V) flux and (VI) other sources.  
The rightmost column shows the total fractional systematic uncertainty due to all sources.
}
\label{tab:systematics}
\end{table}
\endgroup

The vertex energy distribution is sensitive to the multiplicity of low energy charged hadrons in the final state. 
Systematic uncertainties on this distribution are evaluated with the same methods 
used for the cross-section measurement.  The largest uncertainties in the
distribution come from the detector's response to protons (constrained by test beam measurements~\cite{minerva_nim}), the Birk's law constant discussed above, and GENIE's final state interactions model.  The latter is evaluated by varying the underlying model tuning parameters within their systematic uncertainties.



\begin{figure}[tp]
\centering
\includegraphics[width=\columnwidth]{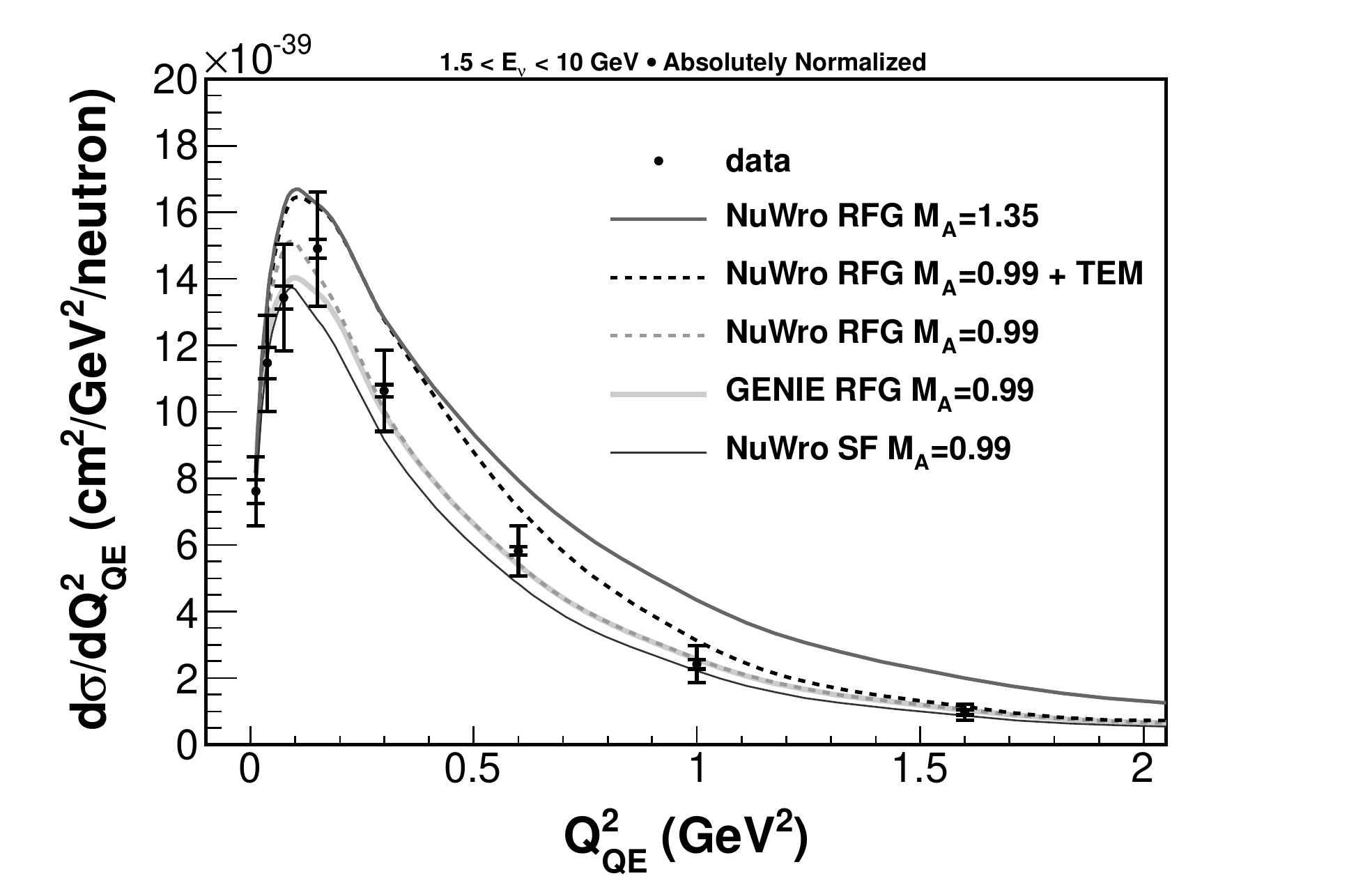}
\caption{Neutrino quasi-elastic cross-section as a function of $Q^2_{QE}$ compared with several different models of the interaction.}
\label{fig:xsec_q2}
\end{figure}

\begin{figure}[tp]
\centering
\includegraphics[width=\columnwidth]{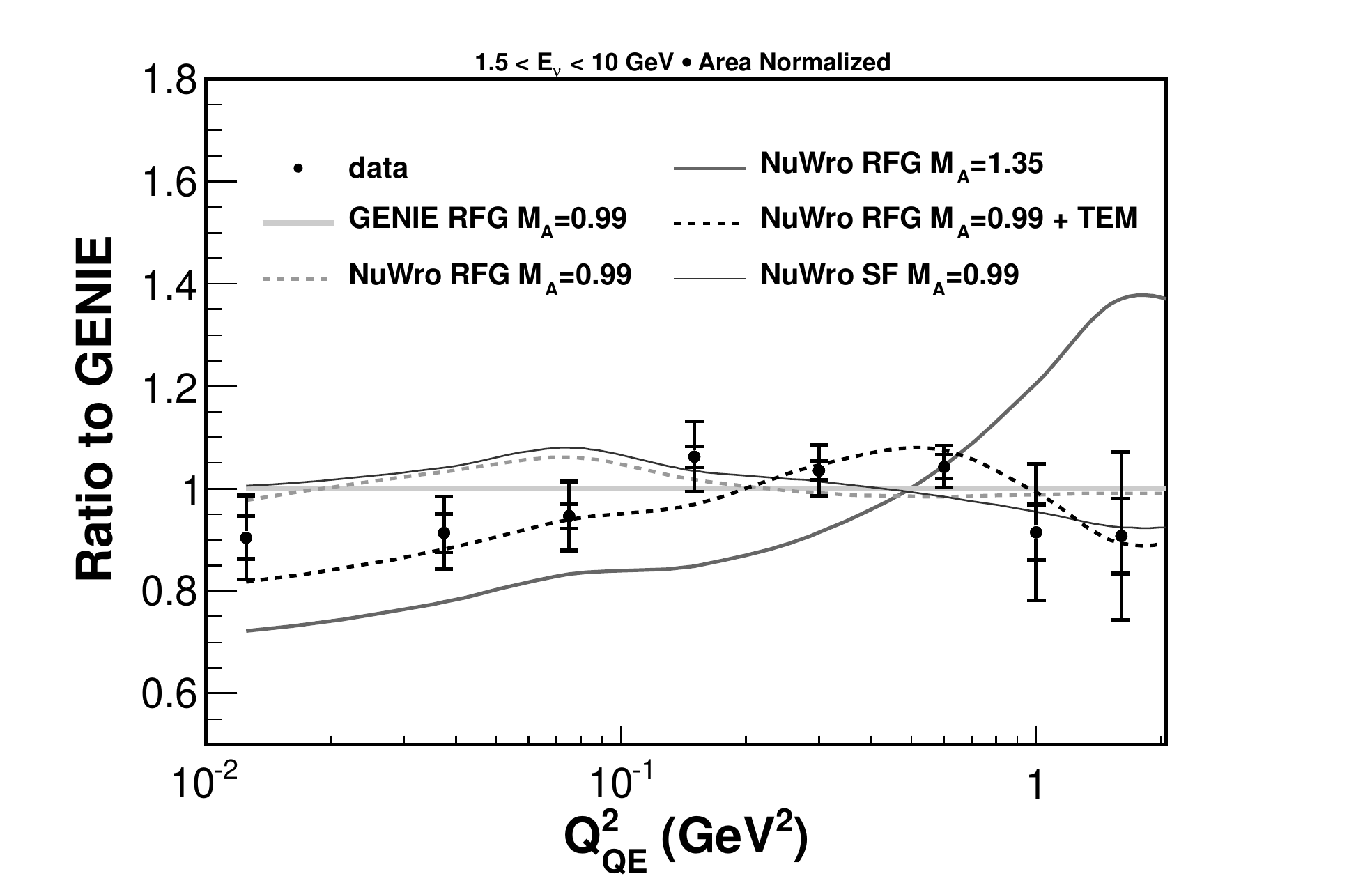} 
\caption{ \label{fig:xsec_q2_shape_ratio} Ratio between the measured neutrino $d\sigma/dQ^2_{QE}$ shape in $Q^2_{QE}$ and several different models, where the denominator is the GENIE default quasi-elastic cross-section.}
\end{figure}

The measured differential cross-section $d\sigma/dQ^2_{QE}$ is shown in 
Table~\ref{tab:xsec} and Fig.~\ref{fig:xsec_q2}. Integrating over the flux from 1.5 to 10~GeV, we find\footnoterecall{footsupp}
  $\sigma=0.93 \pm 0.01 \mbox{(stat)} \pm 0.11 \mbox{(syst)} \times \unit[10^{-38}]{cm^2/neutron}$.  Figures~\ref{fig:xsec_q2} and \ref{fig:xsec_q2_shape_ratio} and Table~\ref{tab:chi2} compare the data to the RFG model in the GENIE event generator and a 
set of calculations made with the NuWro generator~\cite{Golan:2012wx}.  

\begingroup
\squeezetable
\begin{table}
\begin{tabular}{ccc}
$Q^2_{QE}$ & Cross-section  &  Fraction of \\
{(GeV$^2$) }    & {($10^{-38}\mathrm{cm}^2/\mathrm{GeV}^2/$neutron) } &  Cross-section {(\%)} \\ \hline
$0.0 - 0.025$ & $0.761\pm0.035\pm0.097$ &  $2.15\pm0.10\pm0.17$ \\ 
$0.025 - 0.05$ & $1.146\pm0.047\pm0.137$ &  $3.24\pm0.13\pm0.22$ \\ 
$0.05 - 0.1$ & $1.343\pm0.034\pm0.156$ &  $7.60\pm0.19\pm0.50$ \\ 
$0.1 - 0.2$ & $1.490\pm0.028\pm0.170$ &  $16.85\pm0.32\pm1.04$ \\ 
$0.2 - 0.4$ & $1.063\pm0.019\pm0.120$ &  $24.06\pm0.43\pm1.06$ \\ 
$0.4 - 0.8$ & $0.582\pm0.013\pm0.074$ &  $26.33\pm0.58\pm0.85$ \\ 
$0.8 - 1.2$ & $0.242\pm0.014\pm0.053$ &  $10.95\pm0.64\pm1.45$ \\ 
$1.2 - 2.0$ & $0.097\pm0.008\pm0.024$ &  $8.81\pm0.71\pm1.43$ \\ 
\hline
\end{tabular}
\caption{Flux-averaged differential cross-sections and the fraction of the cross-section in bins of $Q^2_{QE}$.  In each measurement, the first error is statistical and the second is systematic.}
\label{tab:xsec}
\end{table}
\endgroup


Different models of nuclear effects in quasi-elastic
scattering lead to significant variations in the shape of $d\sigma/dQ^2$ from
the expectation of the RFG model.
In particular, correlations between nucleons not considered in the mean field RFG approach are predicted to contribute to the cross-section at neutrino energies below 2~GeV~\cite{Martini:2010ex,Martini:2013sha,Nieves:2005rq}.
Figure \ref{fig:xsec_q2_shape_ratio} compares the shape of the measured cross section to five different models of the quasi-elastic process on carbon.  
The GENIE prediction, based on a RFG nuclear model and dipole axial form factor with $M_A = $ 0.99 \GeV, is taken as a reference; the data and other models are normalized to have the same total cross section across the range shown before forming the ratio.  The NuWro calculations utilize an axial-vector form factor parameterized with a dipole form that has one free parameter,  
the axial mass $M_A$, and also incorporate different
 corrections for the nuclear medium. 
There is little sensitivity 
to replacement of the Fermi gas with a spectral function (SF) 
model of the target nucleon energy-momentum relationship~\cite{Benhar:1994hw}.  
The neutrino data are marginally more compatible, at least in $Q^2_{QE}$ shape, with a higher axial
mass extracted from fits of the MiniBooNE neutrino quasi-elastic data
in the RFG model ($M_A=\unit[1.35]{GeV/c^2}$)~\cite{AguilarArevalo:2007ab} than with that extracted from deuterium data ($M_A=\unit[0.99]{GeV/c^2}$). 
As with the corresponding antineutrino results~\cite{nubarprl}, our data are in best agreement with  a transverse enhancement model (TEM)
 with $M_A=\unit[0.99]{GeV/c^2}$.  This model implements an enhancement of the magnetic form factors of bound nucleons that has been extracted from electron-carbon scattering data~\cite{Bodek:2011ps}, and is the only one of this type that is applicable at neutrino energies above 2~GeV.  Table \ref{tab:chi2} shows a comparison using $\chi^2$ values between the measured cross section and the five NuWro models considered.  

\begingroup 
\squeezetable 
\begin{table} 
\begin{tabular}{c|cccc}
NuWro  &  ~RFG~ & ~RFG~  & ~RFG~ & ~SF~  \\  
Model   &          & +TEM~  & &  \\ \hline
$M_A$ (\GeVcc) &  0.99 & 0.99 & 1.35 & 0.99  \\  \hline \hline
Rate $\chi^2$/d.o.f. & 3.5 & 2.4 & 3.7 & 2.8 \\
Shape $\chi^2$/d.o.f. & 4.1 & 1.7 & 2.1 & 3.8 \\
\end{tabular}
\caption{Comparisons between the measured $d\sigma/dQ^2_{QE}$ (or its shape in $Q^2_{QE}$) and different models implemented using the NuWro neutrino event generator, expressed as $\chi^2$ per degree of freedom (d.o.f.) for eight (seven) degrees of freedom.  The $\chi^2$ computation in the table accounts for significant 
correlations between the data points caused by systematic uncertainties. }
\label{tab:chi2}
\end{table}
\endgroup

\begin{figure}[tp]
\centering
\includegraphics[width=0.90\columnwidth]{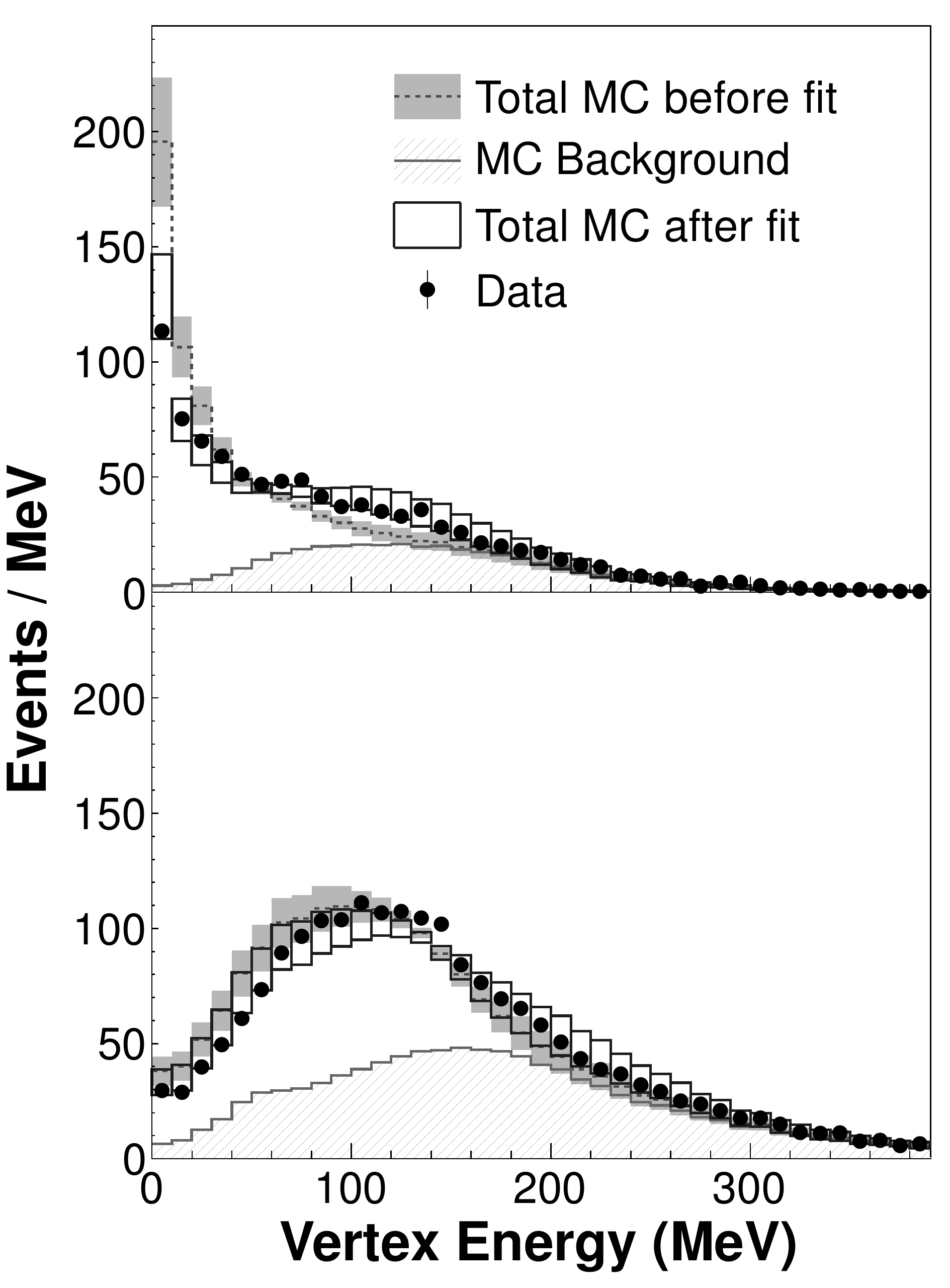}
\caption{ Reconstructed vertex energy of events passing the selection criteria in the data (points with statistical errors) compared to the GENIE RFG model (shown with systematic errors) for $Q^2_{QE} < 0.2$ GeV$^2/$c$^2$ (top) and for $Q^2_{QE} > 0.2$ GeV$^2/$c$^2$ (bottom).
}
    \vspace{-10pt}
\label{fig:vtx_eng}
\end{figure}

Experience from electron quasi-elastic scattering on carbon suggests that
multibody final states are dominated by initial-state $np$
pairs~\cite{Donnelly:1999sw,Martini:2009uj,Subedi:2008zz}.  This could lead to an expectation of final state $pp$
pairs in neutrino quasi-elastic scattering and $nn$ pairs in the
analogous antineutrino channel. The vertex energy measurement, shown
in Fig.~\ref{fig:vtx_eng}, is sensitive to these effects.  These data
prefer the addition of a final state proton with less than 225~MeV
kinetic energy in 25$\pm$1(stat)$\pm$9(syst)\% of the events.  The
corresponding result in the antineutrino mode~\cite{nubarprl}, in
contrast, prefers the {\em removal} of a final state proton in
10$\pm$1(stat)$\pm$7(syst)\% of the events. 
The systematic uncertainties for the two samples 
are positively correlated with a correlation
coefficient of $+0.7$, implying that the observed difference is unlikely 
to be due to one of the systematic uncertainties considered.  
The systematic uncertainties are primarily from the detector response to
protons and uncertainties in reactions in the target nucleus that
absorb or create final state protons.  Independent of models, elastic
and inelastic nucleon reactions which might produce additional final
state protons in the neutrino data should have analogous reactions in
the anti-neutrino data, and the difference in the two results makes it
unlikely that any modification of final state nucleon interactions can
explain the discrepancy.  Pion FSI processes, especially absorption,
would produce more protons in the neutrino reaction and neutrons in
the antineutrino reaction, but the associated uncertainties are
included in the total systematic errors.  
The observed patterns in the
neutrino and antineutrino channels, combined with the observation that
electron quasi-elastic scattering with multinucleon final states in
carbon produces primarily final state $np$ pairs, suggests an initial
state of strongly correlated $np$ pairs also may participate in the
neutrino quasi-elastic interaction.


\ifnum\sizecheck=0
  \begin{acknowledgments}

This work was supported by the Fermi National Accelerator Laboratory
under US Department of Energy contract
No. DE-AC02-07CH11359 which included the \minerva construction project.
Construction support also
was granted by the United States National Science Foundation under
Award PHY-0619727 and by the University of Rochester. Support for
participating scientists was provided by NSF and DOE (USA) by CAPES
and CNPq (Brazil), by CoNaCyT (Mexico), by CONICYT (Chile), by
CONCYTEC, DGI-PUCP and IDI/IGI-UNI (Peru), by Latin American Center for
Physics (CLAF) and by RAS and the Russian Ministry of Education and Science (Russia).  We
thank the MINOS Collaboration for use of its
near detector data. Finally, we thank the staff of
Fermilab for support of the beamline and detector.

\end{acknowledgments}

  \bibliographystyle{apsrev4-1}
\bibliography{CCQE-nu}

\fi

\ifnum\PRLsupp=0
  \clearpage
  \newcommand{\qsq}{\ensuremath{Q^2_{QE}}\xspace}
\renewcommand{\textfraction}{0.05}
\renewcommand{\topfraction}{0.95}
\renewcommand{\bottomfraction}{0.95}
\renewcommand{\floatpagefraction}{0.95}
\renewcommand{\dblfloatpagefraction}{0.95}
\renewcommand{\dbltopfraction}{0.95}
\setcounter{totalnumber}{5}
\setcounter{bottomnumber}{3}
\setcounter{topnumber}{3}
\setcounter{dbltopnumber}{3}

\begingroup
\squeezetable
\begin{table*}[!h]
{\normalsize \appendix{Appendix: Supplementary Material}\hfill\vspace*{4ex}}
\tabcolsep=0.11cm
\begin{tabular}{c|cccccccc}
\hline
$\qsq$ (GeV$^2$) Bins & $0.0 - 0.025$ & $0.025 - 0.05$ & $0.05 - 0.1$ & $0.1 - 0.2$ & $0.2 - 0.4$ & $0.4 - 0.8$ & $0.8 - 1.2$ & $1.2 - 2.0$ \\ 
\hline
Cross-section in bin 
 & 0.761 & 1.146 & 1.343 & 1.490 & 1.063 & 0.582 & 0.242 & 0.097 \\ 
($10^{-38}\mathrm{cm}^2/\mathrm{GeV}^2/$neutron)  & $\pm$ 0.104 & $\pm$ 0.144 & $\pm$ 0.160 & $\pm$ 0.172 & $\pm$ 0.122 & $\pm$ 0.075 & $\pm$ 0.055 & $\pm$ 0.025\\
\hline
$Q^2_{QE}$ (GeV$^2$) & & & & & & & & \\
$0.0 - 0.025$  & 1.000 & 0.869 & 0.882 & 0.873 & 0.832 & 0.690 & 0.415 & 0.327 \\ 
$0.025 - 0.05$  &  & 1.000 & 0.905 & 0.917 & 0.882 & 0.727 & 0.457 & 0.357 \\ 
$0.05 - 0.1$  &  &  & 1.000 & 0.945 & 0.928 & 0.751 & 0.455 & 0.356 \\ 
$0.1 - 0.2$  &  &  &  & 1.000 & 0.946 & 0.788 & 0.481 & 0.385 \\ 
$0.2 - 0.4$  &  &  &  &  & 1.000 & 0.865 & 0.600 & 0.514 \\ 
$0.4 - 0.8$  &  &  &  &  &  & 1.000 & 0.756 & 0.741 \\ 
$0.8 - 1.2$  &  &  &  &  &  &  & 1.000 & 0.888 \\ 
$1.2 - 2.0$  &  &  &  &  &  &  &  & 1.000 \\ 
\hline
\end{tabular}
\caption{The measurement of the neutrino differential cross-sections in $Q^2_{QE}$,
their total (statistical and systematic) uncertainties, and the correlation matrix for these uncertainties}
\end{table*}
\endgroup

\begingroup
\squeezetable
\begin{table*}[!h]
\tabcolsep=0.11cm
\begin{tabular}{c|cccccccc}
\hline
$\qsq$ (GeV$^2$) Bins & $0.0 - 0.025$ & $0.025 - 0.05$ & $0.05 - 0.1$ & $0.1 - 0.2$ & $0.2 - 0.4$ & $0.4 - 0.8$ & $0.8 - 1.2$ & $1.2 - 2.0$ \\ 
\hline
\% of Cross-section  & 
2.15	& 3.24	& 7.60	& 16.85	& 24.06	& 26.33	& 10.95	& 8.81	\\
in bin  &$\pm	0.20	$&$\pm	0.26	$&$\pm	0.53	$&$\pm	1.09	$&$\pm	1.14	$&$\pm	1.03	$&$\pm	1.58	$&$\pm	1.60$	\\
\hline
$Q^2_{QE}$ (GeV$^2$) & & & & & & & & \\
$0.0 - 0.025$  & 1.000 & 0.689 & 0.712 & 0.684 & 0.557 & -0.175 & -0.585 & -0.623 \\ 
$0.025 - 0.05$  &  & 1.000 & 0.745 & 0.770 & 0.653 & -0.211 & -0.631 & -0.694 \\ 
$0.05 - 0.1$  &  &  & 1.000 & 0.840 & 0.793 & -0.212 & -0.736 & -0.787 \\ 
$0.1 - 0.2$  &  &  &  & 1.000 & 0.817 & -0.173 & -0.780 & -0.825 \\ 
$0.2 - 0.4$  &  &  &  &  & 1.000 & -0.129 & -0.752 & -0.795 \\ 
$0.4 - 0.8$  &  &  &  &  &  & 1.000 & -0.142 & 0.060 \\ 
$0.8 - 1.2$  &  &  &  &  &  &  & 1.000 & 0.760 \\ 
$1.2 - 2.0$  &  &  &  &  &  &  &  & 1.000 \\ 
\hline
\end{tabular}
\caption{The measurement of the {\em shape} of the neutrino differential cross-sections for $Q^2_{QE}<2.0$~GeV$^2$, their total (statistical and systematic) uncertainties, and the correlation matrix for these uncertainties}
\end{table*}
\endgroup

\begingroup
\squeezetable
\begin{table*}[b]
\begin{tabular}{l|ccccccccccccc}
\hline 
$E_\nu$ in Bin & 
$1.5 - 2$ &
$2 - 2.5$ &
$2.5 - 3$ &
$3 - 3.5$ &
$3.5 - 4$ &
$4 - 4.5$ &
$4.5 - 5$ &
$5 - 5.5$ \\
$\nu_\mu$ Flux
(neutrinos/cm$^2$/POT ($\times 10^{-8}$) &
$0.310	$ &
$0.409	$&
$0.504	$&
$0.526   $&
$0.423	$&
$0.253	$&
$0.137	$&
$0.081	$ \\
\hline
$E_\nu$ in Bin & 
$5.5 - 6$ &
$6 - 6.5$ &
$6.5 - 7$ &
$7 - 7.5$ &
$7.5 - 8$ &
$8 - 8.5$ &
$8.5 - 9$ &
$9 - 9.5$ &
$9.5 - 10$ \\
$\nu_\mu$ Flux
(neutrinos/cm$^2$/POT ($\times 10^{-8}$) &
$0.055   $&
$0.043	$&
$0.036	$&
$0.031	$&
$0.027	$&
$0.024	$&
$0.021	$&
$0.019	$&
$0.017	$ \\ \hline
\end{tabular}
\caption{The calculated muon neutrino flux per proton on target (POT) for the data included in this analysis}
\end{table*}
\endgroup

\fi

\end{document}